\documentclass[10pt,twocolumn,letterpaper]{article}

\usepackage[pagenumbers]{cvpr} 

\usepackage{microtype}
\newcommand{\inlineheading}[1]{\vspace{3pt}\noindent\textbf{{#1}.}}

\usepackage{tikz}
\usepackage{color}
\usepackage{colortbl}

\definecolor{mygray}{gray}{.9}

\definecolor{cvprblue}{rgb}{0.21,0.49,0.74}
\usepackage[pagebackref,breaklinks,colorlinks,allcolors=cvprblue]{hyperref}

\def\confName{CVPR}
\def\confYear{2025}

\title{SoundVista: Novel-View Ambient Sound Synthesis via Visual-Acoustic Binding}

\author{Mingfei Chen\textsuperscript{1,2$^{*}$}\qquad 
Israel D. Gebru\textsuperscript{2}\qquad 
Ishwarya Ananthabhotla\textsuperscript{3} \qquad
Christian Richardt\textsuperscript{2} \vspace{1.2mm}\\
Dejan Markovic\textsuperscript{2} \qquad 
Jake Sandakly\textsuperscript{2} \qquad 
Steven Krenn\textsuperscript{2} \qquad 
Todd Keebler\textsuperscript{2} 
\vspace{1.2mm}\\
Eli Shlizerman\textsuperscript{1} \qquad
Alexander Richard\textsuperscript{2}
\vspace{2.5mm}\\
\textsuperscript{1}University of Washington\quad \textsuperscript{2}Codec Avatars Lab, Pittsburgh, Meta \quad
\textsuperscript{3}Reality Labs Research, Meta
\vspace{1.25mm}\\
}

\begin{document}

\twocolumn[{
\renewcommand\twocolumn[1][]{#1}
\maketitle
\begin{center}
\vspace{-0.5em}
    \includegraphics[width=2.0\columnwidth,keepaspectratio]{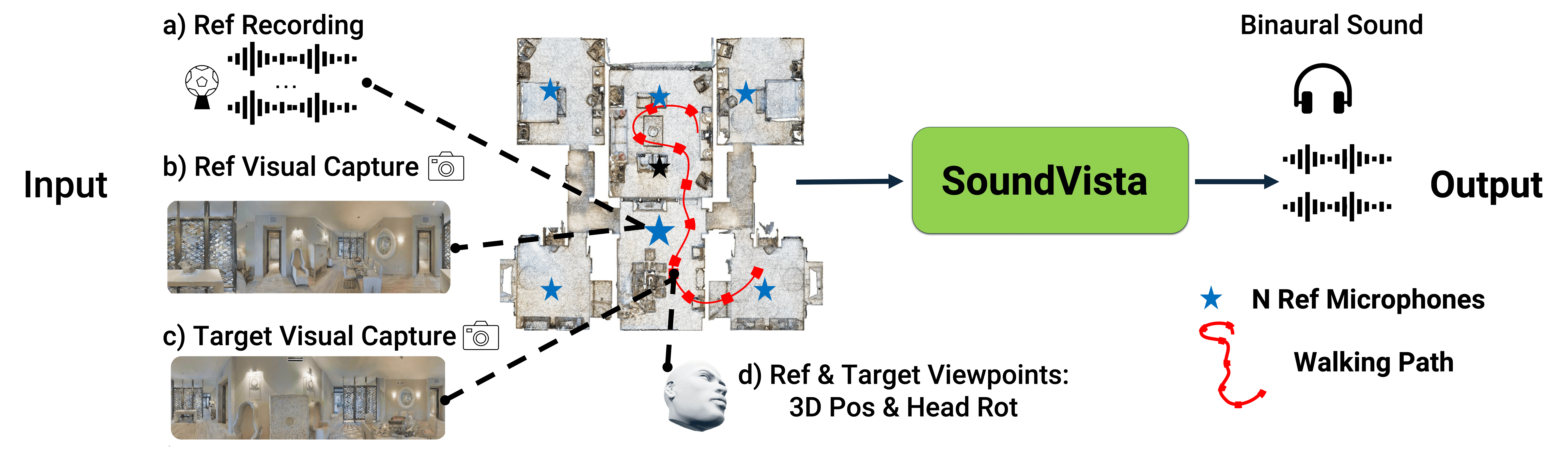}
    \captionof{figure}{\textbf{SoundVista}: a novel method that synthesizes binaural ambient sound for arbitrary scenes from novel viewpoints. 
    Our method leverages pre-acquired audio recordings and visual data captured from sparsely distributed reference points and synthesizes binaural audio consistent with the target 3D position and pose.
    }\label{fig:teaser}
\end{center}
}]

{\let\thefootnote\relax\footnotetext{{* Work done during an internship at Meta.}}}

\begin{abstract}
We introduce SoundVista, a method to generate the ambient sound of an arbitrary scene at novel viewpoints. Given a pre-acquired recording of the scene from sparsely distributed microphones, SoundVista can synthesize the sound of that scene from an unseen target viewpoint. The method learns the underlying acoustic transfer function that relates the signals acquired at the distributed microphones to the signal at the target viewpoint, using a limited number of known recordings. Unlike existing works, our method does not require constraints or prior knowledge of sound source details. Moreover, our method efficiently adapts to diverse room layouts, reference microphone configurations and unseen environments. To enable this, we introduce a visual-acoustic binding module that learns visual embeddings linked with local acoustic properties from panoramic RGB and depth data. We first leverage these embeddings to optimize the placement of reference microphones in any given scene. During synthesis, we leverage multiple embeddings extracted from reference locations to get adaptive weights for their contribution, conditioned on target viewpoint. We benchmark the task on both publicly available data and real-world settings. We demonstrate significant improvements over existing methods.

\end{abstract}

\newcommand{\mf}[1]{\textcolor{cyan}{#1}}
\vspace{-1.5em}
\section{Introduction}
\label{sec:intro}

Recent advances in 3D reconstruction and Novel-View Synthesis~(NVS) have significantly enhanced our ability to create photorealistic visual models of real-world scenes~\cite{VRNeRF,wu2022scalable, jang2022egocentric,ramakrishnan2021habitat}. These developments have paved the way for applications in 3D virtual tours, experience recreation, and spatial media. However, the audio counterpart -- Novel-View Acoustic Synthesis (NVAS) \cite{chen2023novel,ahn2023novel} -- has been under-explored compared to its visual counterpart and has not received the rigorous attention required to generate acoustic virtual scenes that match their visual surroundings. 

To address this gap, we introduce \textit{SoundVista}: a novel method that creates a truly immersive acoustic experience. \textit{SoundVista} can generate realistic and spatially accurate binaural audio at novel viewpoints in arbitrary scenes given a sparse set of sample recordings from different viewpoint.

Unlike novel-view synthesis for visual 3D scenes, which are mostly static, NVAS faces significant challenges due to the dynamic nature of real-world acoustic environments. Ambient sound -- the overall acoustic state describing all sounds in a scene \cite{zhang2018ambient} -- can encompass multiple time-varying, non-stationary sound sources without limitations on the type and number of sound sources. 
In an ideal scenario, with complete information about the ambient sound (\ie clean signals of individual sound sources and their locations over time), and a fully reconstructed 3D visual model, one could synthesize the ambient sound at any location using standard acoustic renderers with room impulse responses (RIRs) \cite{ahn2023novel,chen2020soundspaces, chen2024real,meshrir,su2022inras}.
In real-world scenes, however, we lack detailed information about the sound sources that comprise the ambient sound. Accurately determining their clean source content and emitter positions is a challenging problem. 

NVAS also faces challenges in balancing data sampling costs and synthesis quality. Current methods in the acoustic community use grids of microphones, and techniques ranging from interpolation \cite{mroz2021production,olgun2023sound} to complex signal processing \cite{10445940,borra2019soundfield} to synthesize audio at novel locations. However, these approaches do not scale well to large spaces, and efficient sampling and data utilization remain open questions.

Recent deep learning based NVAS methods \cite{chen2023novel,chen2023everywhere,fewshotir,vam,liang23avnerf} utilize multimodal data, including visual inputs, to transfer sparse reference sounds to binaural sound at target viewpoints. These methods often simplify the task by focusing on re-synthesizing primary sounds such as speech and music, with a limited number of sound sources (typically 1–2). They often ignore other sounds that contribute to the natural characteristics of the ambient sound scene. While feasible with few references, they struggle to adapt to diverse, large scenes with complex layout and acoustic environment. Furthermore, the lack of an optimal reference location sampling strategy hinders adaptation, leading to a loss of acoustic details and making it difficult to synthesize high-quality sound that has accurate binaural effects for target viewpoints.

In this paper, we propose a novel approach that avoids the challenge of obtaining granular information about individual sound sources and addresses some of the shortcomings of existing NVAS techniques. Our method relies on sparsely distributed ``reference microphones" to capture acoustic snapshots, taking their recorded audio signals as ``reference recordings". This allows us to get a holistic representation of the acoustic environment from references recordings. Given many examples of reference recordings, and sounds recorded at known locations within the scene, we formulate NVAS as learning the transformation from reference recordings to sound recorded at target viewpoints. At inference time, given any pre-acquired reference recordings of the scene as input and a query for an arbitrary target position, we expect the model to output binaural audio that is acoustically consistent with the query viewpoint and the content from reference recordings.

To address the limitations of existing NVAS techniques in reference microphone placement, we develop a multi-modal approach for optimal reference location sampling. Our approach can adapt to diverse, large complex scenes using a Visual-Acoustic Binding (VAB) module. 
VAB learns acoustically relevant features (VAB embeddings) by pretraining an encoder to align visual features from panoramic RGB-D captures with acoustic features from echo responses. Using VAB embeddings from numerous candidate reference locations, we employ a sampler that automatically identifies representative spots by finding spatial regions with similar embeddings. These spatial regions represent areas with similar acoustic properties and are usually free from obstacles that significantly hinder sound propagation. This approach optimizes reference microphone placement within a limited reference budget which enhances the overall performance.

Furthermore, to manage varying numbers of references -- which typically depend on scene size and budget -- and their unequal contributions to the final audio, we introduce an adaptive reference integration module. This module models sequences via a transformer, using VAB features to reweight reference inputs and viewpoints. These reweighted inputs then serve as conditions for the final binaural audio renderer. The resulting conditions are both scene-adapted and content-invariant, enabling effective handling of various reference microphone configurations and audio content.

To demonstrate the effectiveness of our approach, we benchmark the proposed model in both real-world setting and in a challenging simulated dataset derived from Matterport3D \cite{Matterport3D} scenes with SoundSpaces \cite{chen2020soundspaces}. SoundVista outperforms existing methods~\cite{ahn2023novel,chen2023novel,liang23avnerf,chen2023everywhere,vam} on scenes with multiple varieties of sound sources, handling varying numbers up to ten sound sources.
\vspace{-0.25em}

\section{Related Work}
\label{sec:related_works}
\vspace{-0.25em}

\inlineheading{Acoustic Scene Synthesis}
Traditional methods focus on estimating the room impulse response (RIR) to recreate spatial audio. This is achieved by convolving the sound from each emitter with the RIR corresponding to the emitter-listener pair and summing the results \cite{ratnarajah2020ir, ratnarajah2021fast,richard2022deep, luo2021learning, su2022inras,Liang2023NeuralAC,chen2024real,Ratnarajah2023AVRIRAR, fewshotir}. 
These RIR-based techniques often require detailed source information, such as a clean signal of each source and its precise location, which are typically unknown in real-world scenarios, making them challenging to implement.
Alternative approaches use scene-descriptor images for direct spatial sound synthesis based on reference sound input, such as visual acoustic matching \cite{vam, chen2023av_dereverb} or visual-guided audio spatialization \cite{mono2bi,2.5D,xu2021visually, Apnet}.
However, these methods may be inaccurate when viewpoints change continuously.

\inlineheading{Novel-view Acoustic Synthesis}
\begin{figure*}[t]
    \centering
    \includegraphics[width=0.98\linewidth]{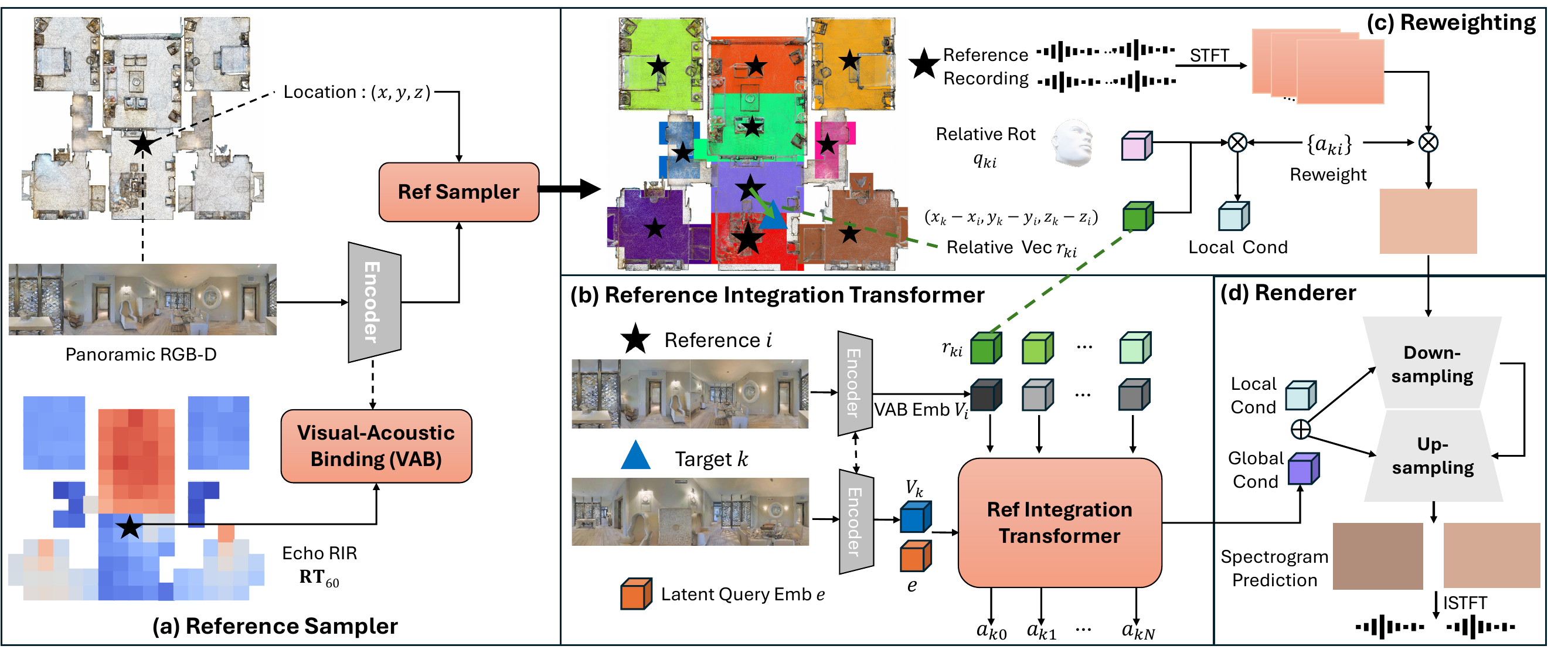}
    \captionsetup{aboveskip=-0.25mm}
    \caption{\textbf{Details of the SoundVista Pipeline:}
    (a)~The reference location sampler selects optimal reference locations leveraging embeddings from visual-acoustic binding~(VAB).
    (b)~The reference integration transformer uses VAB embeddings to derive contribution weights for each reference.
    (c)~Reweighting by contribution weights adjusts and integrates reference recording channels and pose conditioning for precise sound synthesis.
    (d)~The spatial audio renderer converts reweighted channels and conditions to binaural sound at the target viewpoint.}
        \label{fig:main}
        \vspace{-3mm}
\end{figure*}
Novel-view acoustic synthesis techniques address limitations in acoustic scene synthesis by learning transfer functions from reference sounds to target viewpoints \cite{chen2023novel,liang23avnerf, chen2023everywhere,warpnet,wavenet}. Methods such as \textit{Few-shotRIR}~\cite{fewshotir} and \textit{BEE}~\cite{chen2023everywhere} use multiple references but lack optimal reference sampling strategies, relying on heuristic settings such as a single close reference \cite{chen2023novel,liang23avnerf,wavenet,warpnet}, random sampling \cite{fewshotir}, or fixed references \cite{chen2023everywhere}, which limit adaptation to diverse scene layouts.  
In contrast, our method optimizes reference location sampling via visual-acoustic binding and derives adaptive reference contribution weights, enhancing sound synthesis accuracy and generalization.

\inlineheading{Acoustic-Visual Learning}
Techniques such as CLIP~\cite{radford2021learning, xue2022clip,luo2021clip4clip,fang2021clip2video,lin2022frozen} have successfully aligned visual and linguistic inputs to learn matched deep semantic representations. These advances led to recent works to leverage complementary nature of audio and visual data for acoustic related tasks. For example, 
AV-RIR~\cite{Ratnarajah2023AVRIRAR} binds RIR with panoramic images to improve the late reverberation modeling of RIR. Advances in multi-modal visual understanding have facilitated  sound source localization \cite{hu2020discriminative,jiang2022egocentric,tao2021someone,mo2023audio}, audio-visual speech enhancement and source separation \cite{ephrat2018looking, owens2018audio, michelsanti2021overview, zhou2019vision,Tian2021CyclicCO,ye2024lavss}, as well as acoustic scene reconstruction \cite{fewshotir,vam,singh2021image2reverb,Ratnarajah2023AVRIRAR,chen2023novel,chen2023everywhere}. These works underscore the strong link between visual and acoustic modalities. Inspired by these approaches, we developed a visual encoder that aligns features from RGB-D visual data with acoustic features from echo responses. The alignment enables us to infer acoustic properties using visual data only. As a result, our method does not require knowing acoustic parameters or RIRs during inference. It is also easily adaptable to various scene and acoustic environments. 
\looseness-1
\section{Method} 
\label{sec:method}
\subsection{Problem Formulation}
\label{sec:problem_formulation}
Given $N$ pairs of co-located microphones and cameras, we aim to strategically place them in the acoustic scene. Let $\mathbf{L}_i=(x_i, y_i, z_i) \in \mathbb{R}^3 $ denote the location of the devices and $\boldsymbol\theta_i \in \mathbb{R}^2$ denote their orientations. The microphones capture reference audio recordings $\mathbf{A}_i$, and the cameras capture panoramic RGB-D views $\mathbf{V}_i$. These ``reference captures" encapsulate the ambient sound and visual properties, providing the basis for reconstructing binaural ambient sound $\Tilde{\mathbf{A}}_k$ for a target listener at arbitrary location $\mathbf{L}_k=(x_k, y_k, z_k) \in \mathbb{R}^3$ with orientation $\boldsymbol\theta_k \in \mathbb{R}^2$. 
The reference recording can be of any type, but we use ambisonic microphones as they are ideal for capturing sound fields from multiple directions and simplify binaural signal extraction.
In contrast to existing methods \cite{ahn2023novel}, we avoid enumerating sound sources. Instead, we learn a transfer function $\mathcal{F}$ from reference captures to the target as:
\vspace{-2mm}
\begin{align}
\mathcal{F} &: \left(\big\{\mathbf{A}_i, g_i, {r}_{ki}, \boldsymbol\theta_i\big\}_{i=1}^{N}, g_k, \boldsymbol\theta_k \right) \mapsto \Tilde{\mathbf{A}}_k \text{,} \\
g_i &= \phi(\mathbf{V}_i),~ g_k = \phi(\mathbf{V}_k), \text{and} ~ {r}_{ki} = \mathbf{L}_k - \mathbf{L}_i,  \nonumber
\vspace{-5mm}
\end{align}
\noindent where $g_i$ and $g_k$ are Visual-Acoustic Binding (VAB) embeddings extracted with a pre-trained visual encoder $\phi(\cdot)$, designed to align visual with 
acoustic representations. ${r}_{ki}$ is relative vector introduced to avoid overfitting on absolute locations and to handle diverse scene coordinate systems. 

We parameterize $\mathcal{F}$ using neural networks that incorporate a transformer network and spatial audio renderer module. The transformer generates an adaptive mask to weight the contribution of each reference recording based on the target viewpoint and the VAB embeddings associated with reference viewpoints. We call this the ``Reference Integration Transformer". The spatial audio renderer then processes the reweighted audio channels along with conditional information from the target view features to produce the final binaural sound. We provide an overview of these modules in \cref{fig:main}, with details explained in the following sections.

\subsection{Visual-Acoustic Binding (VAB)}
\label{sec:VAB}
Obtaining acoustic properties of real-world scenes is challenging \cite{chen2024real}. We propose to leverage RGB-D data, which can provide rich information linked to acoustic properties of a scene. For example, depth data provides information about obstacles and room geometry, while pixel-level textures reveal material differences and detailed obstacle information. 

The goal is to infer the acoustic properties of a scene using only the visual data. To achieve this, we first collected extensive paired panoramic RGB-D and acoustic echo responses data by navigating walkable areas in the SoundSpaces simulator \cite{chen2020soundspaces}. We then trained a neural network, which we refer as ``Visual-Acoustic Binding (VAB) module", to predict the acoustic representations from the visual features. 

\inlineheading{Acoustic Representation}
In acoustics, an impulse response is a function of time and the positions of the emitter and the listener \cite{vigran2014building}.
It describes how sound propagates through a medium and interacts with the environment \cite{raghuvanshi2014parametric, raghuvanshi2018parametric}. 
To simplify data collection and infer local acoustic characteristic, we focus on echo impulse responses \cite{fewshotir}, where the emitter and target locations are the same. 
We extracted reverberation time (aka $\text{RT}_{60}$ parameter) from echo response to use as acoustic representation.  $\text{RT}_{60}$ measures how long it takes for the acoustic energy to decay by $60$dB and can reveal information about room geometry, obstacles and reflection. \Cref{fig:main}(a) shows an example $\text{RT}_{60}$ map of a scene. Significant value changes in $\text{RT}_{60}$ map can indicate major obstacles or surface variations, highlighting key acoustic regions.

\inlineheading{Visual Representation}
Following prior works \cite{chen2023novel,Ratnarajah2023AVRIRAR}, we use panoramic RGB-D captures as inputs and extract visual features from ResNet-18 \cite{he2016deep}. We train the VAB module to predict $\text{RT}_{60}$. Therefore, effective binding visual representation with acoustic representation.

\subsection{Reference Location Sampler} \label{sec:ref_loc_sample}
While having more microphones and cameras is ideal for this task, in practice resources are often limited. To maximize performance and make the best use of available resources, we propose a strategic approach for placement of reference microphones throughout a given scene.  

We argue that ideal microphone placements align with the acoustic partitions of the scene--areas with unique acoustic properties and free from obstacles such as walls. To identify these partitions, we capture panoramic RGB-D images from all candidate locations. One can use novel-view synthesis \cite{VRNeRF} to render these images; without needing actual photographs at each spot. We then extract VAB embeddings from each capture. VAB embeddings which correspond to scene acoustic parameters serve as strong cues for identifying acoustic partitions, such as distinguishing regions separated by obstacles. To enhance reliability, we combine the extracted VAB embeddings with positional information, allowing them to complement each other. Using these embeddings, we perform data clustering of the candidate locations and take the center of each cluster as a reference location.  This is illustrated in \cref{fig:main}(a).

\vspace{-0.25em}
\subsection{Reference Integration Transformer} \label{sec:ref_att}
We want our model to work effectively across diverse scenes. This requires the model to adapt to varying numbers of reference audio and visual inputs. Logically, larger scenes would benefit from more microphones and cameras, while smaller spaces can operate efficiently with fewer resources. Moreover, since the task (\cref{sec:problem_formulation}) is to transfer reference recordings to target sounds at specific viewpoints, distant microphones, or those in non-adjacent acoustic partitions usually contribute significantly less to the process than closer ones. However, weighting based solely on distance is insufficient due to potential obstacles such as walls and objects. 

To address these issues, we propose a transformer network. We treat each reference input as part of a sequence, allowing us to manage varying numbers of reference inputs. To derive the weights for their unequal contributions to the final audio, we exploit VAB embeddings extracted from visual captures at the reference and target viewpoints. 

We formed the {queries} input by combining the VAB embedding at the target location $g_k$, with a learnable latent query embedding $\mathbf{e}$. The query is initialized from a normal distribution and optimized during training, to adjust the contribution of reference microphones based on their relative location vector to the target location. Let $g_{k}^{e}=[g_k~||~\mathbf{e}] \in \mathbb{R}^{\text{C}}$ to denote the {queries}. The {keys} and {values} inputs are created by combining the VAB embedding of each reference with its relative vector ${r}_{ki}$. Let $g_{i}^{r}=[g_i~||~r_{ki}] \in \mathbb{R}^\text{C}$ to denote the keys and values of i\textsuperscript{th} reference for the attention operation. The attention weight $a_{ki}$ can be calculated as follows:
\begin{align}
\vspace{-3.0em}
\label{transformer}
a_{ki} = \frac{g_{k}^{e} \cdot {g_{i}^{r}}\top}{\sqrt{\text{C}}},  \quad i=\{1, 2, \ldots, N\}.
\end{align}
The attention weight $a_{ki}$ indicates contribution of the reference $i$ to the target $k$ normalized by the Softmax function over $N$ references. Higher weights indicate greater influence.

\subsection{Spatial Audio Renderer} \label{sec:renderer}
The Spatial Audio Renderer takes the reference recordings and conditional information to generate binaural sound at the target viewpoint. 

\inlineheading{Reweighting Reference Recordings}
We apply the Short-Time Fourier Transform (STFT) to the reference recordings to extract their spectrograms. Then, we reweight the channels of each reference recording using the attention weights \(a_{ki}\) from \cref{sec:ref_att} to integrate them, encoding the result into a latent space as the input for the renderer.

\inlineheading{Reweighting Condition Inputs}
The binaural sound is derived based on the target's position and pose. To capture the different aspects of the binaural effects, we decouple the conditioning into global and local components. This accounts for the cause of spatialization effects at a given target point, which may primarily vary from the distance to the sound source or remain invariant to head orientation. The global condition $c_g$ determines how the target viewpoint's relative position to reference locations influences the binaural sound. The local condition $c_l$ accounts more for the effects of head orientation. They are defined as follows:
\begin{align}
\vspace{-2.0em}
&\phantom{={}} c_g = \sum^{N}_{i=1}{a_{ki} \cdot \psi_1(g_i \, \| \, {r}_{ki})} \text{,} \label{eq:global_cond} \\
&\phantom{={}} c_l = \left( \sum^{N}_{i=1} a_{ki} \cdot \psi_2(r_{ki} \, \| \, {q}_{ki}) \right) + \sigma(\theta_k). \label{eq:local_cond}
\end{align}
Here \( {q}_{ki} \) is the relative orientation quaternion for target \( k \) relative to reference \( i \); \(\psi_1\) and \(\psi_2\) are MLP layers projecting concatenated inputs to a latent condition embedding; and \(\sigma(\mathbf{\theta}_k)\) denotes the target's rotational features from \(\mathbf{\theta}_k\).

\inlineheading{Renderer}
The Audio Renderer network is designed as a stacked U-Net \cite{ronneberger2015u,song2021scorebased}, depicted in \cref{fig:main}(d). It features down-sampling and up-sampling blocks that incorporate input conditions at each layer. To preserve quality and content details from the reference input, skip connections are employed at multiple resolutions. After processing the integrated reference recordings, we apply the inverse STFT to convert the output spectrogram into a binaural waveform.

Training from scratch in complex environments with challenging audio content can hinder the renderer to adapt to head orientation while producing high-quality audio. To address this, we pretrain the binauralization capability of our renderer by fixing the target location and varying head rotations and audio sources. After pre-training, the renderer effectively understands binauralization across different orientations, enhancing the accuracy of spatial effect modeling.

\subsection{Loss}
We design the loss function as a weighted combination of three components, each regulating different aspects of the model:
1) \textit{Waveform Loss} measures the mean squared error between the predicted and target waveforms, ensuring precision of the predicted waveform amplitude and phases.
2) \textit{Binaural Interaural Level Difference (ILD) Loss} \cite{ick2024spatially} focuses on the energy difference between binaural audio channels to ensure accurate spatial effect prediction.
3) \textit{Multi-resolution Spectrogram Magnitude Loss} \cite{yamamoto2020parallel} ensures that the predicted audio matches the target spectrogram magnitude across multiple resolutions. It includes two terms: the first compares log-scaled magnitudes, and the second, the Scaled Spectrogram Magnitude Loss, calculates magnitude distance over the target scale, addressing high variance in audio magnitude. We found that resolution with large FFT and small hop sizes can benefit modeling ambient noise, such as air vibrations.

\section{Experiments}

\subsection{Benchmarks and Metrics}

\inlineheading{N2S Benchmark} \label{sec: n2s_benchmark}
Captured in a real room of 8.5 by 6 meters dimensions with two internal rooms, the N2S benchmark utilizes microphone arrays with 32 microphones for 32-channel recordings. 
Visual captures are rendered by the NeRF-based 3D NVS model, VR-NeRF~\cite{VRNeRF}, after it has been trained on visual data collected by navigating the room. 11 static microphone arrays are uniformly distributed to provide reference ambient recordings, while 6 mobile arrays cover 557 locations with 15 seconds of recording per location. Motion capture system, OptiTrack, records 6DoF tracking data for both reference and target microphones. 35\% of these locations with 20 orientations are selected for training. 

\inlineheading{Soundspace-Ambient Matterport3D}
We build this benchmark based on Matterport3D scenes, comprising 39 complex training scenes and 23 unseen scenes. It is challenging as each scene can include 2 to 10 sound sources, selected from 128 different sounds, from fan noise to speech. Reference locations are spaced every 5 meters on average, with simulated second-order ambisonics RIRs to render reference sounds by Soundspaces~\cite{chen2020soundspaces}. 85\% of walkable locations are used for training with azimuth angles of 0, 90, 180, and 270.

\inlineheading{Metrics}
We evaluate sound synthesis performance using four key metrics \textbf{(the lower, the better)}: 1) L1 distance of STFT spectrogram (\textbf{STFT})~\cite{mono2bi} for left and right channels; 2) Magnitude Spectrogram Distance (\textbf{MAG})~\cite{mono2bi}, measuring the closeness of reconstructed audio to the groundtruth; 3) Energy Envelope Error (\textbf{ENV})~\cite{mono2bi}, assessing the Euclidean distance between energy envelopes of groundtruth and predicted audio channels; 4) Left-Right Energy Ratio Error (\textbf{LRE})~\cite{chen2023novel}, evaluating binaural effect accuracy by calculating energy ratio difference between left and right channels.

\begin{table*}[tb]
\caption{\label{tab:mp3d_main}%
    \textbf{Results Comparison on Soundspace-Ambient Benchmark:}
    Average metrics for 10,189 samples in 39 seen scenes with novel target locations and sources, and 6,534 samples in 23 unseen scenes.
    Sampling strategies: \textit{location} only, \textit{vis+location} (our sampler), and \textit{w/o VAB} (ours without VAB pretraining).
    \textit{SoundVista} Ref Num $k$: only references with top-$k$ contribution weights are used for fair comparison.}
\small
\centering
\vspace{-3mm}
\begin{tabular}{|llc|cccc|cccc|}
\hline
&     &  &   \multicolumn{4}{c|}{Seen Scenes} &  \multicolumn{4}{c|}{Unseen Scenes} \\
Method      & Sampling   &Ref Num       & STFT $\downarrow$                   & MAG $\downarrow$                                 & ENV $\downarrow$                    & LRE $\downarrow$                     & STFT  $\downarrow$                   & MAG $\downarrow$                                            & ENV $\downarrow$                    & LRE $\downarrow$ \\
\hline
nearest GT     & location  &1 &4.448 &0.351	&0.154	&1.596 & 4.034 & 0.353    & 0.155       &1.617                \\
nearest GT     & w/o VAB &1 &4.414 &0.341	&0.151	&1.576 &4.680 &0.347	&0.153	&1.537	  \\
nearest GT     & vis+location &1 	&3.916 &0.336	&0.146	&1.572       & 3.835 & 0.344   & 0.151    & 1.557   \\

interp GT      & location  &4   &3.922   &0.326		&0.144	&1.584  & 3.410  & 0.327     & 0.145   & 1.587 \\
interp GT     &w/o VAB  &4	&3.660 &0.319	&0.142	&1.570  &3.766 &0.320	&0.142	&1.531  \\
interp GT      & vis+location &4 &3.179  & 0.313		&0.137	&1.559  & 3.415 & 0.321   & 0.141    & 1.531   \\
\hline
AV-NeRF~\cite{liang23avnerf} & vis+location &1 &9.424 &0.426 &0.195 &1.922 &9.321 &0.428 &0.196 &1.979\\
VAM~\cite{vam}  & vis+location &1  &5.224 &0.420 &0.178	&1.902 &4.936 &0.436 &0.182	&1.977\\
ViGAS~\cite{chen2023novel} & vis+location   &1   &3.740 &0.361 &0.154	&2.040 &3.438 &0.371		&0.157	&2.051 \\
\cellcolor{mygray}SoundVista (Ours)   &\cellcolor{mygray}vis+location &\cellcolor{mygray}1 &\cellcolor{mygray}2.526	&\cellcolor{mygray}0.291		&\cellcolor{mygray}0.132	&\cellcolor{mygray}1.408	&\cellcolor{mygray}2.676	&\cellcolor{mygray}0.309 &\cellcolor{mygray}0.140	&\cellcolor{mygray}1.386 \\ 
BEE~\cite{chen2023everywhere}& vis+location  &4  &4.098 &0.365 &0.162 &2.083 &5.635 &0.396 &0.178 &2.131\\
\cellcolor{mygray}SoundVista (Ours)   &\cellcolor{mygray}vis+location &\cellcolor{mygray}4 &\cellcolor{mygray}2.444	&\cellcolor{mygray}0.289		&\cellcolor{mygray}0.130	&\cellcolor{mygray}1.390	&\cellcolor{mygray}2.517	&\cellcolor{mygray}0.305 &\cellcolor{mygray}0.137	&\cellcolor{mygray}\bf1.371 \\ 
Few-shotRIR~\cite{fewshotir}   & vis+location &all     &5.937 &0.459 &0.213	&1.892 &5.457 &0.471 &0.215	&1.960 \\
\cellcolor{mygray}SoundVista w/o vis &\cellcolor{mygray}location &\cellcolor{mygray}all   &\cellcolor{mygray}3.228 &\cellcolor{mygray}0.306  &\cellcolor{mygray}0.141 &\cellcolor{mygray}1.425   &\cellcolor{mygray}2.890  &\cellcolor{mygray}0.312  &\cellcolor{mygray}0.142  &\cellcolor{mygray}1.439                \\ 
\cellcolor{mygray}SoundVista (Ours)   &\cellcolor{mygray}vis+location &\cellcolor{mygray}all &\cellcolor{mygray}\bf 2.442	&\cellcolor{mygray}\bf 0.289		&\cellcolor{mygray}\bf 0.130	&\cellcolor{mygray}\bf 1.390	&\cellcolor{mygray}\bf 2.514	&\cellcolor{mygray}\bf 0.305 &\cellcolor{mygray}\bf 0.137	&\cellcolor{mygray}1.372 \\

\hline
\end{tabular}
\vspace{-3mm}
\end{table*}

\begin{table}[htb]
\caption{\label{tab:n2s}%
    \textbf{Testing Results Comparison on N2S Benchmark.}
    In this real-world scene, \textit{SoundVista} with visual modality largely boosts the performance accuracy, especially on the binaural effect (LRE).}
\small
\centering
\vspace{-3mm}
\begin{tabular}{|l|cccc|}
\hline
Method  & STFT $\downarrow$   & MAG $\downarrow$ & ENV $\downarrow$   & LRE $\downarrow$   \\
\hline
Nearest DSP &2.420  &0.212		&0.136	&1.447 \\
Interp DSP  &1.659 &0.203	&0.142	&1.383 \\
AV-NeRF~\cite{liang23avnerf} &2.194 &0.187 &0.119 &0.840\\
Few-shotRIR~\cite{fewshotir} &1.765  &0.199	&0.134	&0.909 \\
VAM~\cite{vam}  &1.972 &0.190	&0.119	&0.916 \\		
BEE~\cite{chen2023everywhere} &1.471 &0.200	&0.141	&0.995\\
ViGAS~\cite{chen2023novel}  &1.201 &0.185	&0.119	&0.873 \\
\cellcolor{mygray}SoundVista w/o vis  &\cellcolor{mygray}1.242 &\cellcolor{mygray}0.185	&\cellcolor{mygray}0.118	&\cellcolor{mygray}0.894 \\
\cellcolor{mygray}SoundVista (Ours)  &\cellcolor{mygray}\bf 1.073 &\cellcolor{mygray}\bf 0.177	&\cellcolor{mygray}\bf 0.113	&\cellcolor{mygray}\bf 0.776 \\
\hline
\end{tabular}
\vspace{-5mm}
\end{table}

\subsection{Comparison with Baselines}
We compare our method with the following baseline approaches:
(1) \textbf{Nearest GT}: Binaural sound from the nearest reference microphone aligned with the target orientation.
(2) \textbf{Interp GT}: Employs binaural sound at the target orientation from the four nearest reference microphones, interpolating based on distance to the target location.
(3) \textbf{AV-NeRF}~\cite{liang23avnerf}: A NeRF-based system that synthesizes binaural audio from a given camera pose and RGB-D renderings. We adapted it to use recordings, poses, and visual context from references while preserving its core model components, enabling fair comparisons in dynamic scenes.
(4) \textbf{Few-shotRIR}~\cite{fewshotir}: A transformer-based method that infers RIRs from a sparse set of observed images and echoes. Adapted to replace the reference and target impulse responses with the ambient sounds at the corresponding viewpoints.
(5) \textbf{VAM}~\cite{vam}: Matches the acoustics of input audio with a target image.
(6) \textbf{BEE}~\cite{chen2023everywhere}: A generalizable rendering pipeline that reconstructs the binaural audio at an arbitrary listener location using inputs from sparse reference audio-visual samples in the scene.
(7) \textbf{ViGAS}~\cite{chen2023novel}: Transforms sound to the target viewpoint given the observed audio and visual captures at the reference viewpoint.
Specifically, for \textit{AV-NeRF}, \textit{VAM}, and \textit{ViGAS}, which need a single reference, we use the nearest reference microphone. \textit{BEE} is adapted to use the nearest 4 microphones. Deep-learning baselines requiring visual inputs utilize panoramic RGB-D images and the same visual encoder as our model for consistency.

Our results in \cref{tab:mp3d_main} and \cref{tab:n2s} demonstrate that our method, \textit{SoundVista}, consistently surpasses baselines across diverse novel scenes and real scenarios. Despite the challenge of obtaining binaural groundtruth (GT) for arbitrary target orientations, \textit{SoundVista}  significantly reduces errors across all metrics compared with \textit{nearest GT} and \textit{interp GT}. 
On the challenging Soundspace-Ambient benchmark, with diverse sound sources and complex layouts, most deep-learning baselines underperform compared to non-learning methods due to the need for robust conditioning models.

For a fair comparison, we evaluate \textit{SoundVista} using the top 1 and 4 reference microphones by contribution weight, respectively. 
Compared to learning-based methods with 1 reference, \textit{SoundVista} outperforms the best baseline \textit{ViGAS} by 32.5\% (STFT), 19.4\% (MAG), 14.3\% (ENV), and 31\% (LRE), demonstrating the strength of our conditioned audio renderer.
The nearest microphone may miss critical audio content, while using the top 4 references improves performance, matching that of all microphones since most references are not required due to their sparse distribution. Since contribution weights are independent of audio content, fewer than 4 references can be selected per target location by precomputing contribution weights once per scene.

While \textit{Few-shotRIR} and \textit{BEE} use audio content to integrate references, the diverse sound distribution makes this challenging. During testing, content often falls outside the training distribution, degrading performance. Thus, more references aren't always better; for example, \textit{Few-shotRIR} did not gain an increase in synthesis accuracy and it was even reduced compared to baselines using only one reference.

\begin{figure*}[htb]
\small
    \centering
    \includegraphics[width=1.0\linewidth]{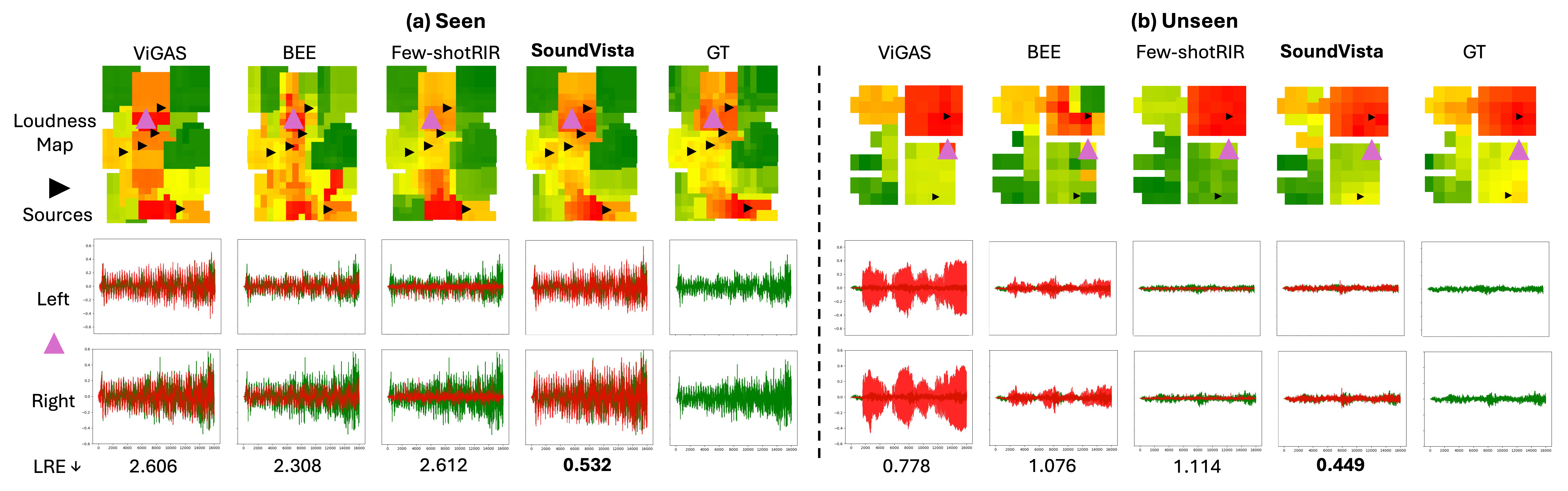}
\captionsetup{aboveskip=-1mm}
\caption{\textbf{Comparison of Qualitative Results:} First row: Loudness Map (black triangle: sources; purple triangle: target viewpoint). Second row: Reconstructed binaural waveform at target viewpoint (red: prediction; green: GT). Third row: LRE (lower for better binaural effect).}
    \label{fig:res_vis_cmp}
    \vspace{-3mm}
\end{figure*}

The deep-learning baselines perform better on the N2S benchmark, which features a single static scene. Compared to these deep-learning baselines, our \textit{SoundVista} still significantly outperforms on accuracy of energy (ENV) and binaural effect (LRE) by 5\% and 7.6\%, respectively.

\subsection{Qualitative Results Comparsion}
\cref{fig:res_vis_cmp} visualizes our results for qualitative comparison with several competitive sound synthesis baselines: \textit{ViGAS}, \textit{BEE}, and \textit{Few-shotRIR}. The first row displays loudness maps, while the second row shows generated binaural waveforms with corresponding LRE error displayed in the third row.
\textit{ViGAS} relies on the nearest microphone, leading to discrete loudness maps in complex layouts and unreliable results when obstacles are present, as shown in the unseen results. Additionally, \textit{ViGAS} struggles with accurate binaural effects shown from the LRE results. \textit{BEE} produces discontinuous loudness maps, deviating from the ground truth. Both \textit{BEE} and \textit{Few-shotRIR} have difficulty generating high-fidelity waveforms close to the groundtruth. Their reliance on incorporating audio content into the renderer's conditioning can bias the renderer leading to confusion and reduced quality when encountering content outside the training distribution. This is particularly common in our complex task setting with diverse audio sources.
In contrast, \textit{SoundVista}, closely matches the GT in loudness maps and binaural waveforms for both seen and unseen scenes, excelling in high-quality novel-view ambient sound synthesis.

\subsection{Ablation Study}

We conducted ablation studies on the Soundspace-Ambient benchmark to verify the contribution of key components.

\inlineheading{Visual-Acoustic Binding (VAB)}
In the design of VAB, we optimized the integration of visual features that correlate with acoustic parameters. \cref{tab:echo_pred} shows the accuracy of $\text{RT}_{60}$ value predictions in novel scenes without finetuning (\textit{w/o finetune}) and with few-shot finetuning at reference locations (\textit{w/ finetune}). We tested various modality inputs: \textit{location} only, panoramic  \textit{rgb}, \textit{depth}, and  \textit{rgb+depth}. Depth alone significantly enhances binding, reducing error by over 50\% compared to use \textit{location} only. With finetuning, \textit{rgb+depth} achieves superior results. Given the ability to capture references in scenes, we select \textit{rgb+depth} as our visual input.

\begin{table}[tb]
\caption{\label{tab:echo_pred}%
    \textbf{$\text{RT}_{60}$ Prediction Results on Matterport3D.}
    \textit{w/ finetune} involves few-shot finetuning on unseen scenes.
    \textit{err}: distance between predicted and GT $\text{RT}_{60}$, \textit{error ratio}: scaled \textit{err} over GT.}
\small
\centering
 \vspace{-3mm}
\begin{tabular}{|l|cc|cc|}
\hline
&     \multicolumn{2}{c|}{w/o finetune} &  \multicolumn{2}{c|}{w/ finetune} \\
Modality   & err $\downarrow$ & err ratio $\downarrow$  & err $\downarrow$ & err ratio $\downarrow$ \\
\hline
location & 0.170      & 0.411        & 0.067           & 0.199         \\
rgb       & 0.117      & 0.236    & 0.044           & 0.126                 \\
depth     & \bf 0.082      & \bf 0.173    & 0.038           & 0.109  \\
rgb+depth & 0.087      & 0.185   & \bf 0.036           & \bf 0.103               \\
\hline
\end{tabular}
 \vspace{-3mm}
\end{table}

\inlineheading{Reference Sampler via VAB}
The Reference Sampler via Visual-Acoustic Binding (VAB) (\textit{vis+location}) significantly enhances the accuracy of non-learning baselines such as \textit{nearest GT} and \textit{interp GT}, as shown in \cref{tab:mp3d_main}. This method improves predictions of sound magnitude (MAG, ENV) and spectrogram phases (STFT). Without pretrained VAB (\textit{w/o VAB}), these benefits diminish, especially in seen scenes. Results in \cref{tab:mp3d_main}, \ref{tab:n2s}, and \ref{tab:mp3d_ablation} further highlight the critical role of visual input, as its absence increases errors across all metrics, underscoring the importance of VAB with visual information for adapting to diverse scene layouts.

\begin{figure}[tb]
\small
    \centering
    \includegraphics[width=1.0\linewidth]{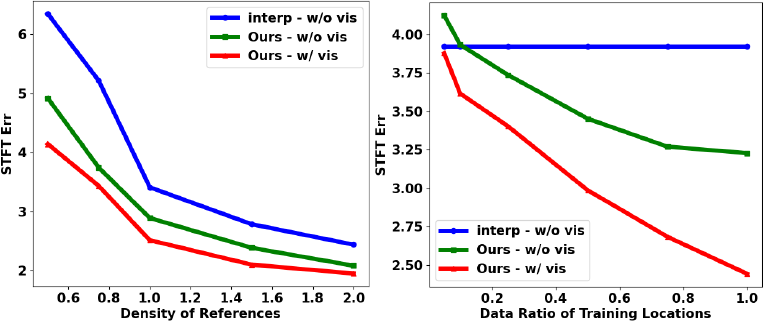}
\caption{\textbf{STFT Error Curve} for different reference densities (left) and training location data ratios (right).}
\captionsetup{aboveskip=-2mm}
    \label{fig:ref_density}
     \vspace{-5mm}
\end{figure}

\noindent \cref{fig:ref_density} (left) illustrates the STFT error curve relative to reference density. Our model is trained with a reference density of 1.0, and tested on novel Soundspace-Ambient scenes with varying densities. Compared to \textit{Ours - w/o vis} and \textit{interp - w/o vis} (\textit{interp GT} with location-only sampler), \textit{Ours - w/ vis} outperforms others, with the performance gap widening as density decreases, demonstrating robustness of ours equipped with VAB vs. reference density variance.

\begin{table}[tb]
\caption{\label{tab:mp3d_ablation}%
    \textbf{Ablations on the SoundSpace-Ambient benchmark.}}
\small
\centering
 \vspace{-3mm}
 
\begin{tabular}{|l|cccc|}
\hline
Method    & STFT $\downarrow$   & MAG $\downarrow$ & ENV $\downarrow$   & LRE $\downarrow$   \\
\hline
full model  &\bf 2.442	&\bf 0.289		&\bf 0.130	&\bf  1.390 \\
w/o vis  &3.228 &0.306  &0.141 &1.425 \\
w/o reweight &2.908  &0.295 &0.134	&1.391\\
w/o render\_pretrain    &2.582 &0.310 &0.135	&1.877 \\ 
w/o decoup\_cond &3.017 &0.300 &0.138 &1.433 \\
\hline
full loss  &2.442	&\bf 0.289		&\bf 0.130	&  1.390 \\
w/o ILD\_loss		&2.462 	&0.292		&0.131		&1.420	\\				
w/o mag\_scale	&\bf 2.424	&0.301		&0.136		&1.422 \\	
w/o wav\_loss &3.422 &0.291 &0.143 &\bf 1.369 \\											
\hline
\end{tabular}
 \vspace{-5mm}
\end{table}

\inlineheading{Reference Integration}
To evaluate the contribution of the adaptive reference integration, we conducted experiments with the \textit{w/o reweight} variant, which uses only the nearest reference microphone along with its corresponding visual and pose information to calculate the condition for the audio renderer. As shown in \cref{tab:mp3d_ablation}, the results of \textit{w/o reweight} compared to full \textit{SoundVista} model reveal a noticeable degradation in performance, particularly in the STFT error, which increases by 19\%. This indicates that the adaptive Reference Integration Transformer and reweighting modules effectively integrate references to enhance sound synthesis accuracy, as the nearest microphone may miss critical content required for the target sound synthesis.

\inlineheading{Conditioned Spatial Audio Renderer}
In the \textit{w/o decoup\_cond} variant, we replaced the decoupled global and local conditions from reweighting in \cref{sec:renderer} with only the global condition plus target rotational features. This absence of local conditions significantly increases STFT error by 23.5\%, MAG error by 3.8\%, and ENV error by 6.2\%.

\noindent As detailed in \cref{sec:renderer}, we pretrain the renderer with binauralization capability to help the model learn the complex task of orientation-sensitive novel view sound synthesis. The \textit{w/o render\_pretrain} variant shows a 35\% increase in LRE error, underscoring the importance of renderer pretraining for achieving a more accurate binaural effect.

\inlineheading{Loss}
The ablations in \cref{tab:mp3d_ablation} highlight the contributions of different loss components. The Waveform Loss (\textit{wav\_loss}) constrains the energy envelope and regulates phase learning. Removing it significantly reduces STFT and ENV accuracy, increasing the STFT error by 29\%. The Scaled Spectrogram Magnitude Loss (\textit{mag\_scale}) calculates magnitude loss over the target magnitude scale, addressing high variance in audio content and mitigating scale differences. Removing \textit{mag\_scale} will increase MAG error largely.

\inlineheading{Robustness w.r.t. Training Data Ratio}
To assess the impact of training location data ratio on performance, we trained two variants, \textit{Ours - w/o vis} and \textit{Ours - w/ vis}, using different proportions of training locations on the Soundspace-Ambient benchmark. \cref{fig:ref_density} (right) shows that STFT error increases as the data ratio decreases. When the data ratio is below 10\%, \textit{Ours - w/o vis} underperforms compared to the non-learning baseline \textit{interp - w/o vis}, while \textit{Ours - w/ vis} maintains superior performance. Additionally, \textit{Ours - w/ vis} benefits more from increased training locations, with STFT error decreasing rapidly as the ratio nears 100\%, widening the performance gap with \textit{Ours - w/o vis} and \textit{interp - w/o vis}.

\begin{figure}[tb]
\small
    \centering
    \includegraphics[width=1.0\linewidth]{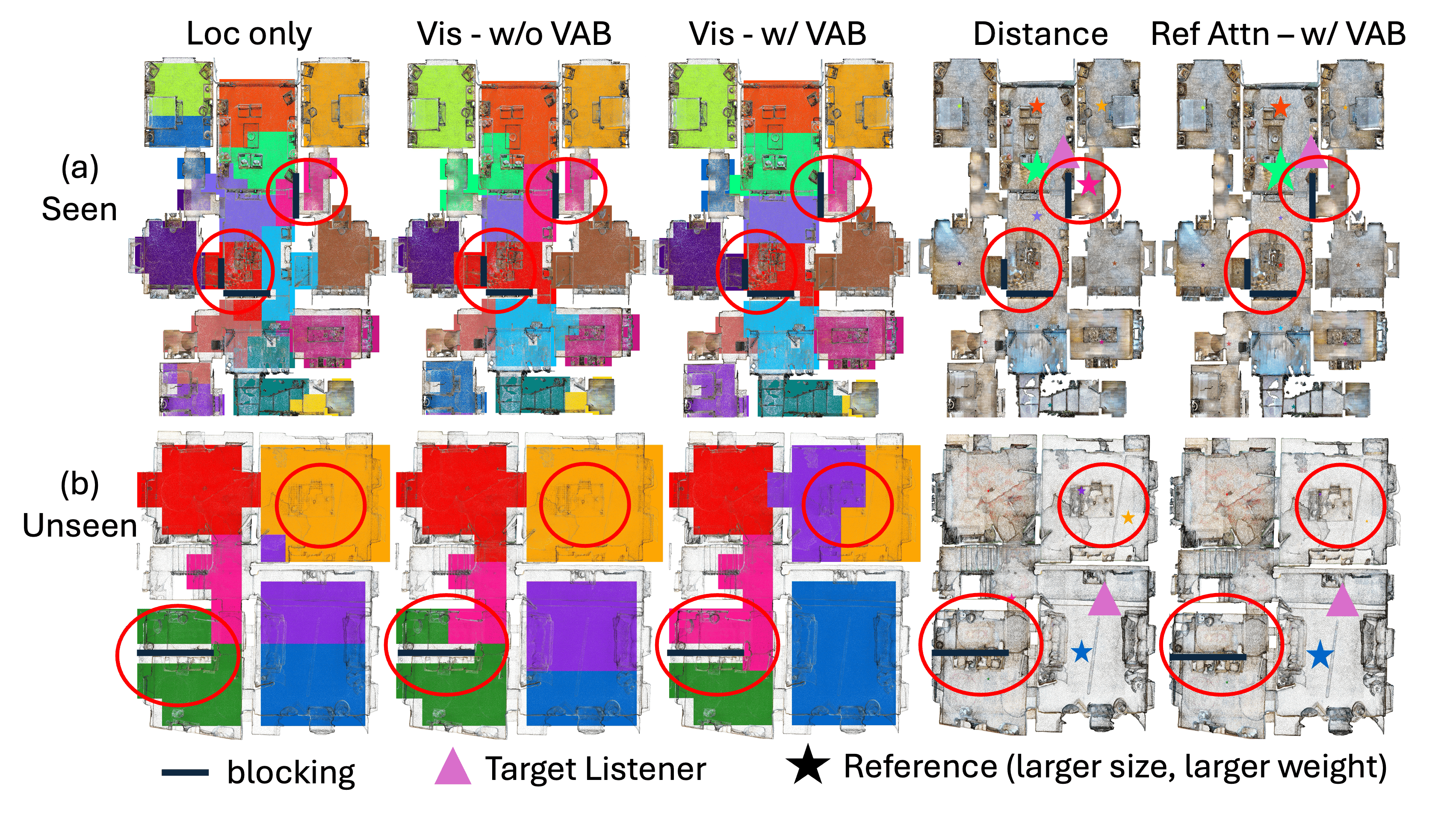}
    \captionsetup{aboveskip=-0.1mm}
    \caption{\textbf{Visualization Analysis.} First three columns: Clustering results for \textit{loc only}, \textit{Vis - w/o VAB} (ours without VAB pretraining), and \textit{Vis - w/ VAB} (ours). Colors indicate cluster regions. VAB helps clustering to align better with room partitions. Last two columns: Attention weights visualization for references, with colored stars (size proportional to weights) and a purple triangle for the target. 
    }
    \label{fig:cluster}
    \vspace{-2mm}
\end{figure}

\subsection{Visualization Analysis}
\cref{fig:cluster} visualizes the clustering results of Reference Sampler via VAB, comparing \textit{loc only}, \textit{Vis - w/o VAB}, and \textit{Vis - w/ VAB}. Different colors represent cluster regions. The Reference Sampler \textit{Vis - w/ VAB} accurately segments clusters aligning with actual room partitions, proving to be more reliable. In contrast, other methods may incorrectly cluster non-adjacent rooms together  which cannot share the same acoustic environment, highlighted by red-circles.

After identifying reference locations, we visualize the attention weights from the Reference Integration Transformer in \cref{fig:cluster} (column 4-5 from left). Colored stars denote sampled references, sized by contribution weight, with the target marked by a purple triangle. Unlike \textit{Distance}-only weights, which may incorrectly prioritize closer references despite obstacles, our model, \textit{Ref Attn - w/ VAB}, trained with the adaptive Reference Integration Transformer, effectively highlights reliable references in similar acoustic environments.
 \vspace{-2mm}
\vspace{-2mm}
\section{Conclusion}
We introduce \textit{SoundVista}, a novel system designed to synthesize ambient sound from arbitrary scenes at novel viewpoints. Our approach introduces a visual-acoustic binding module to effectively learn visual embeddings linked with local acoustic properties. This enables our system to optimize the placement of reference microphones and derive adaptive weights for each microphone's contribution, conditioned on the target viewpoint and visual captures, thereby, facilitating the final conditioned sound synthesis. \textit{SoundVista} adapts to diverse room layouts, microphone configurations, and unseen environments, rendering high-quality binaural ambient sound without requiring prior constraints or detailed knowledge of sound sources. Our results, validated on both publicly available data and real-world settings, demonstrate state-of-the-art sound synthesis accuracy and generalization.

{
    \small
    \bibliographystyle{ieeenat_fullname}
    \bibliography{main}
}

\clearpage
\setcounter{page}{1}
\maketitlesupplementary

\section{Demo Examples}
Please see the attached videos in the supplementary \texttt{demo\_videos} folder. We included videos from Matterport3D scene and our real-world scene (N2S). For the best experience, please \textbf{turn on your audio and use headphones}.

\subsection{Real Scene: N2S Demo}
This demo contain videos from a real-world scene. The scene is captures using $11$ reference microphones, their spatial distribution is shown in \cref{fig:sim2real_cluster} (\textit{Ref Num} = 11). \textbf{Unlike simulated scenes, the real scene presents challenges with diverse natural sounds, including diffuse machine noise and air conditioner vibrations, which are difficult to identify and localize in 3D.} Using reference sounds as input for the Novel-View Ambient Sound Synthesis task proves more effective than attempting to localize and separate sources to render with Room Impulse Responses (RIRs). 

In the demo video (\texttt{0\_0\_n2s\_soundvista.mp4}) of \textit{SoundVista}, three dominant sound sources are clearly identifiable: a TV playing water and bird sounds, a black speaker in the corner playing music, and an air conditioner producing diffuse noise throughout the scene. The sound changes noticeably when entering a small, noisy room with considerable reverberation. As the listener continuously moves in the scene, our model was able to reconstructs these sounds without requiring source counting, localization and RIR data.

\subsection{Soundspace-Ambient Matterport3D Demo}
Videos prefixed with \texttt{1\_x} are results from Matterport3D scenes. We show results from 10 different room that are part of the Soundspace-Ambient benchmark. In \texttt{1\_0\_mp3d\_source\_explain.mp4}, we outline the setup, which include 17 reference points (green stars) and 5 sound sources (blue triangles) distributed throughout the scene. The sources produce various sounds, such as running shower water, engine noise, fireplace crackling, a phone ring, and birds chirping. 

In the videos, the listener (target); shown as a red circle \tikz\draw[red,fill=red] (0,0) circle (.5ex); navigates between rooms throughout the scene. The binaural sound adapts naturally to both viewing orientation and source distance. Though sound transitions remain mostly smooth, crossing between rooms can create more sudden changes because of physical barriers.

\subsection{Comparison with Baselines}
We compare our results with two baselines: \textit{DSP}, a traditional signal processing approach that interpolates binaural sounds from the four nearest reference points using target orientation and distance, and \textit{ViGAS}, a recently proposed learning-based method. \textit{SoundVista} produces better results compared to the baseline methods. Specifically, \textit{SoundVista} smoothly adapts binaural effects to view orientations. 

For example, in the N2S scene, when navigating the TV region (video \texttt{0\_1\_n2s\_comparison\_tvclip.mp4}) and turning around, \textit{DSP} and \textit{ViGAS} fail to properly track the TV sound as it moves from left to front to right. Moreover, \textit{DSP}'s simple interpolation of nearby reference points proves inadequate for handling obstacle effects, resulting in inaccurate sound magnitudes. \textit{ViGAS} introduces sound distortions, especially with bass-heavy music and engine noise, and produces unexpected abrupt changes in sound magnitude. \textit{SoundVista}, in contrast, delivers consistently high-quality (undistored), smooth, and continuous audio. Similar examples are also demonstrated in comparison videos around N2S speaker (video \texttt{0\_2\_n2s\_comparison\_speakerclip.mp4}) and in the Soundspace-Ambient Matterport3D scene  (video \texttt{1\_1\_mp3d\_comparison.mp4}).

\section{Implementation Details}
This section details the implementation of each \textit{SoundVista} module.

\subsection{Visual Acoustic Binding (VAB)}
For training, we partition the panoramic image into four RGB-D views, each of size \({224 \times 224 \times 4}\), and use ResNet-18 as the visual encoder to extract an embedding of dimension 256 for each view. These representations are concatenated as the VAB embedding \( g \) with a dimension of 1024.

\subsection{Reference Location Sampler}~\label{suppl sec: ref sampler}
To determine reference locations within a scene, we first calculate the number of reference points needed by dividing the walkable region's range by a standard distance of 8 meters. With this allocated budget, we then sample locations by clustering all potential walkable reference points.

For each location, we extract the visual representation \( g \) using the pretrained VAB visual encoder. We expand the 3-dimensional location to match the 1024-dimensional \( g \) using sinusoidal encoding and concatenate these representations. We then use K-means clustering to group the candidate locations based on the combined embeddings.

Due to the complexity of Matterport3D scenes with multiple floors, we cluster locations floor by floor. We group walkable locations by height, rounding to the nearest meter. After removing groups with fewer than three locations, we allocate the budget proportionally based on each group's size. This ensures at least one location per group, with groups arranged from smallest to largest to maintain strict budget control. 

After combining all clustering results, we select the walkable location nearest to each floor group's cluster center as the sampled reference location.

\subsection{Reference Integration Transformer}
We deploy a three-layer cross-attention Transformer for reference integration, which features four heads and a dropout ratio of 0.1. The model has a dimension \( C \) of 256 and a feedforward hidden dimension of 512. We use a latent query embedding \( e \) with a dimension of 128. This is concatenated with the projected VAB embedding, which also has a size of 128, to form the queries. The relative vector is encoded using positional encoder with sine-cosine functions, utilizing a frequency number of 10, and is projected to a vector with a dimension of 128.

\subsection{Reweighting}
The dimensions of both the local and global conditions are 256. Specifically, for the local condition, we use sine-cosine functions to embed the rotation quaternion, similar to the approach used in positional encoding.

\subsection{Spatial Audio Renderer}
We utilize the Short-Time Fourier Transform (STFT) to convert waveform audio into the time-frequency domain. The FFT size, window length, and hop length are set to 510, 510, and 128, respectively, and a Hanning window is applied. We chunk the input waveform into segments of length 32641 to form a spectrogram of size \(256 \times 256\).
The renderer consists of a U-Net structure with six downsampling layers and six upsampling layers. The conditions are multiplied to combine with the audio content within the condition layers.

\subsection{Loss and Training}
To balance the loss values, we assign coefficient weights to each of the three loss components: Waveform Loss, Binaural Interaural Level Difference (ILD) Loss, and Multi-resolution Spectrogram Magnitude Loss, with weights of 20, 0.025, and 1.0, respectively. We employ the Adam optimizer for optimization, using an exponentially decaying learning rate starting from \(1 \times 10^{-4}\) over 60 epochs. The batch size is set to 16 for the Soundspace-Ambient benchmark and 24 for the N2S benchmark.
Each batch consists of various training samples from the same scene to optimize memory usage when calculating reference VAB embeddings for reference integration.

\section{VAB for Reference Integration}
\begin{table}[htb]
\centering
\begin{tabular}{|l|cccc|}
\hline
Method    & STFT $\downarrow$   & MAG $\downarrow$ & ENV $\downarrow$   & LRE $\downarrow$   \\
\hline
w/ VAB  &\bf 2.442	&\bf 0.289		&\bf 0.130	&\bf  1.390 \\
w/o VAB &2.580 &0.295	&0.134		&1.403\\
\hline
\end{tabular}
\caption{Ablations for VAB in Reference Integration Transformer.}
\label{tab:vab_ref_attn}
\end{table}

\begin{figure*}[htb]
\small
    \centering
    \includegraphics[width=1.0\linewidth]{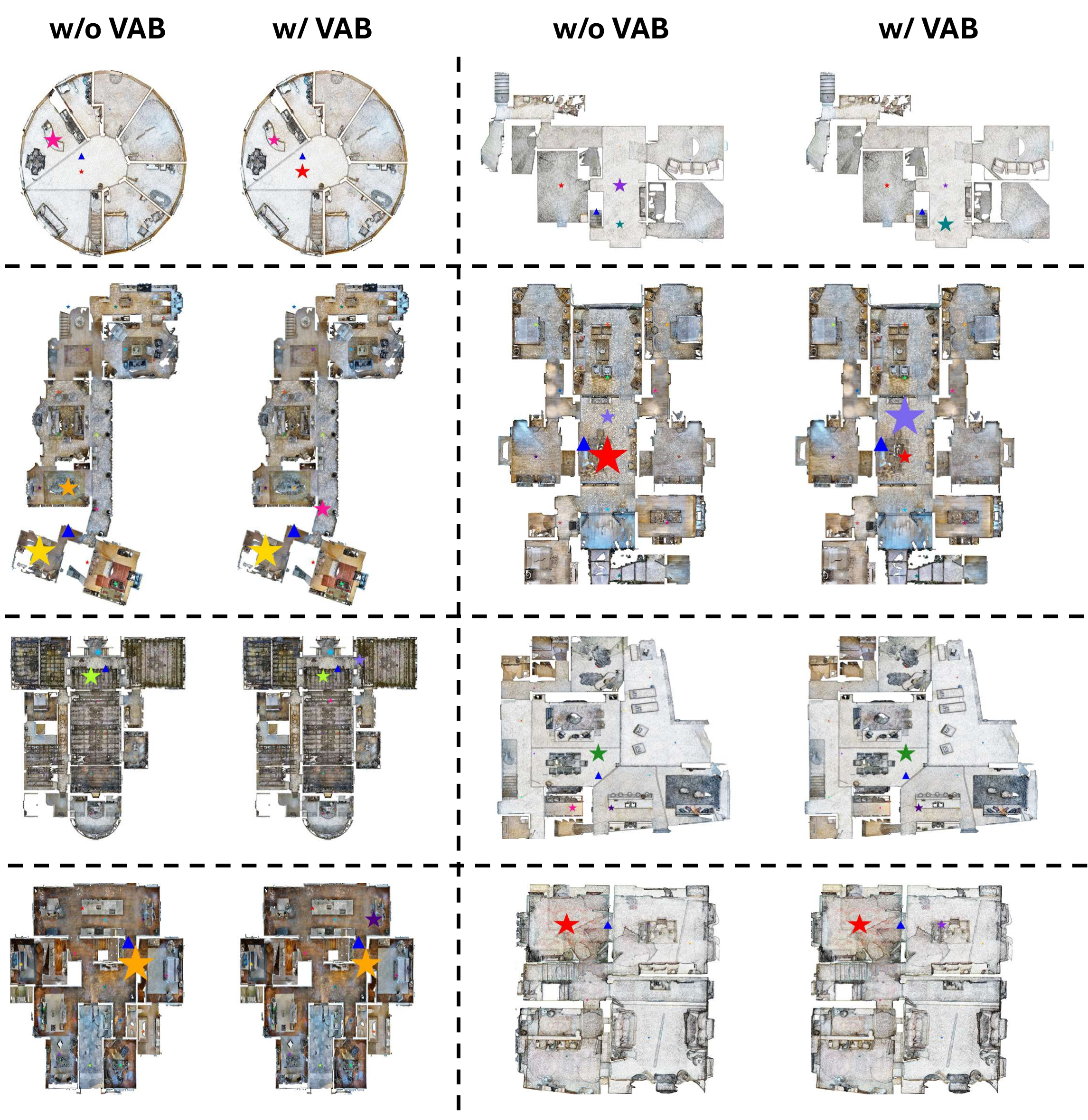}
     \caption{\textbf{Visualization of Reference Contribution Weights.} Colored stars (size proportional to weights) indicate the references and the blue triangle for the target. \textit{w/ VAB} effectively incorporates visual cues to make the contribution weights more reasonable.
    }
    \label{fig:cluster_ref_vab}
\end{figure*}

In this section, we study the effectiveness of using VAB embeddings for the Reference Integration Transformer. We implement a variant that excludes the VAB embeddings from the transformer~(\textit{w/o VAB}) to compare with \textit{SoundVista} with VAB in the transformer~(\textit{w/ VAB)}. We report the ablations results on the Soundspace-Ambient benchmark in \cref{tab:vab_ref_attn} and visualize examples of the reference contribution weights in \cref{fig:cluster_ref_vab}. Compared with \textit{w/o VAB}, \textit{w/ VAB} effectively incorporates visual cues to make the contribution weights more reasonable.

\begin{figure*}[t]
    \centering
    \includegraphics[width=0.95\linewidth]{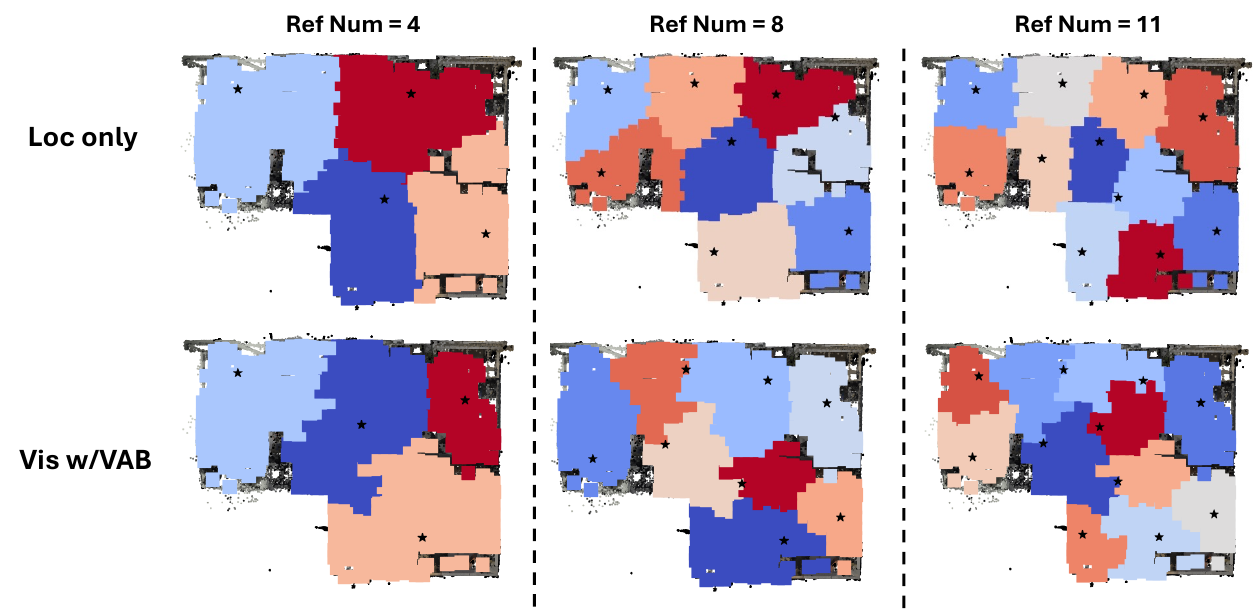}
     \caption{{\textbf{Visualization of Clustering Results on N2S.} Colorized regions are different clusters. The reference location out of existed 11 references that is closest to each cluster center is marked as black star. Our sampler, \textit{Vis w/VAB}, consistently groups locations that are free from obstacles more effectively, demonstrating reliability of VAB from simulated to real scenarios.}
    }
    \label{fig:sim2real_cluster}
\end{figure*}

\section{Extrapolation Performance Analysis}
In our work, the reference microphones are sparsely placed (over 5 meters apart), the edge 
regions of the rooms typically fall outside the convex hull formed by these microphones.
Due to limited in-room data, we cannot track poses or GT sound far beyond the room to evaluate the extrapolation performance. In \cref{fig:res_vis_cmp}, we show the loudness heatmaps for two scenes; while the errors are larger in the edge regions, the results remain reliable. Furthermore, \cref{tab:mp3d_main} demonstrates that using the top selected reference microphone achieves accuracy comparable to using multiple microphones. These findings show SoundVista's ability to extend beyond simple interpolation.

\section{Acoustic Parameter Learning}
We train the acoustic parameter (\(\mathbf{RT}_{60}\)) learning model on walkable locations from 39 ``seen scenes" of Matterport3D in the Soundspace-Ambient benchmark. An MLP is employed to predict the \(\mathbf{RT}_{60}\) value from the VAB embedding \( g \), using L1 loss for supervision. For testing on unseen scenes, we directly use the pretrained visual encoder without fine-tuning for the Novel-View Ambient Sound Synthesis task.

For the \textit{w/ finetune} setting, we aim to study how our acoustic parameter predictor adapts to novel scenes through few-shot learning by finetuning the pretrained prediction model on each of the 23 unseen scenes. Specifically, we uniformly sample the reference locations given the reference budget, maintaining the same average distance as our reference location sampler, but using uniform sampling only. We obtain the \(\mathbf{RT}_{60}\) value as ground truth to supervise the prediction at these locations, which constitutes few-shot fine-tuning on sparsely sampled references. After training per scene, we test the prediction on all walkable locations for each scene separately and average the \(\mathbf{RT}_{60}\) prediction metrics to report accuracy for \textit{w/ finetune}.

\cref{fig:suppl_seen_vis} and \cref{fig:suppl_unseen_vis} illustrate examples of groundtruth and \(\mathbf{RT}_{60}\) predictions for both seen and unseen scenes, respectively. The groundtruth \(\mathbf{RT}_{60}\) map shows that \(\mathbf{RT}_{60}\) values tend to be consistent within a room and are higher in larger spaces without many obstacles, such as open rooms or hallways. This is because sound takes longer to decay in these areas due to fewer reflections or diffusion on surfaces. The \(\mathbf{RT}_{60}\) map is typically discontinuous in regions blocked by obstacles like walls or closed doors.

In scenes seen during training, our predictions closely match the ground truth. For unseen scenes, while the predicted values may deviate in some regions, they can still effectively distinguish different \(\mathbf{RT}_{60}\) areas, accounting for walls and other obstacles that block sound propagation. By applying few-shot finetuning (\textit{w/ finetune}) to correct deviated values, our prediction accuracy can improve significantly.

\section{More Visualization Analysis}
In this section, we present additional visualizations of our clustering results using VAB.

\subsection{Sim2Real Clustering on N2S}
To evaluate the simulate-to-real (sim2real) capability of VAB, which is trained on simulated data from Soundspace, we deploy the pretrained visual encoder in a real N2S room. We cluster the walkable locations using the Reference Sampler (see \cref{suppl sec: ref sampler}) to obtain clusters. 

In \cref{fig:sim2real_cluster}, we visualize the clusters with different reference numbers (\textit{Ref Num} = 4, 8, and 11), coloring each cluster differently. We compare two samplers: \textit{Loc only} and our sampler, \textit{Vis w/VAB}. 
Since the 11 reference locations are already fixed in the real room, we mark the existing reference location closest to the cluster center with black stars, rather than selecting the walkable location nearest to the center. 

\cref{fig:sim2real_cluster} shows that \textit{Loc only} is more likely to incorrectly cluster locations with obstacles in between, especially with fewer reference numbers (4 and 8 compared to 11), making it less effective at identifying obstacles. In contrast, our sampler, \textit{Vis w/VAB}, consistently groups locations that are free from obstacles more effectively, even without any training or supervision in the real scene. This demonstrates the reliability of adapting VAB from simulated to real scenarios.

\subsection{Clustering via VAB}
\begin{figure*}[htb]
\begin{minipage}{1.99\columnwidth}
\small
    \centering
    \includegraphics[width=0.95\linewidth]{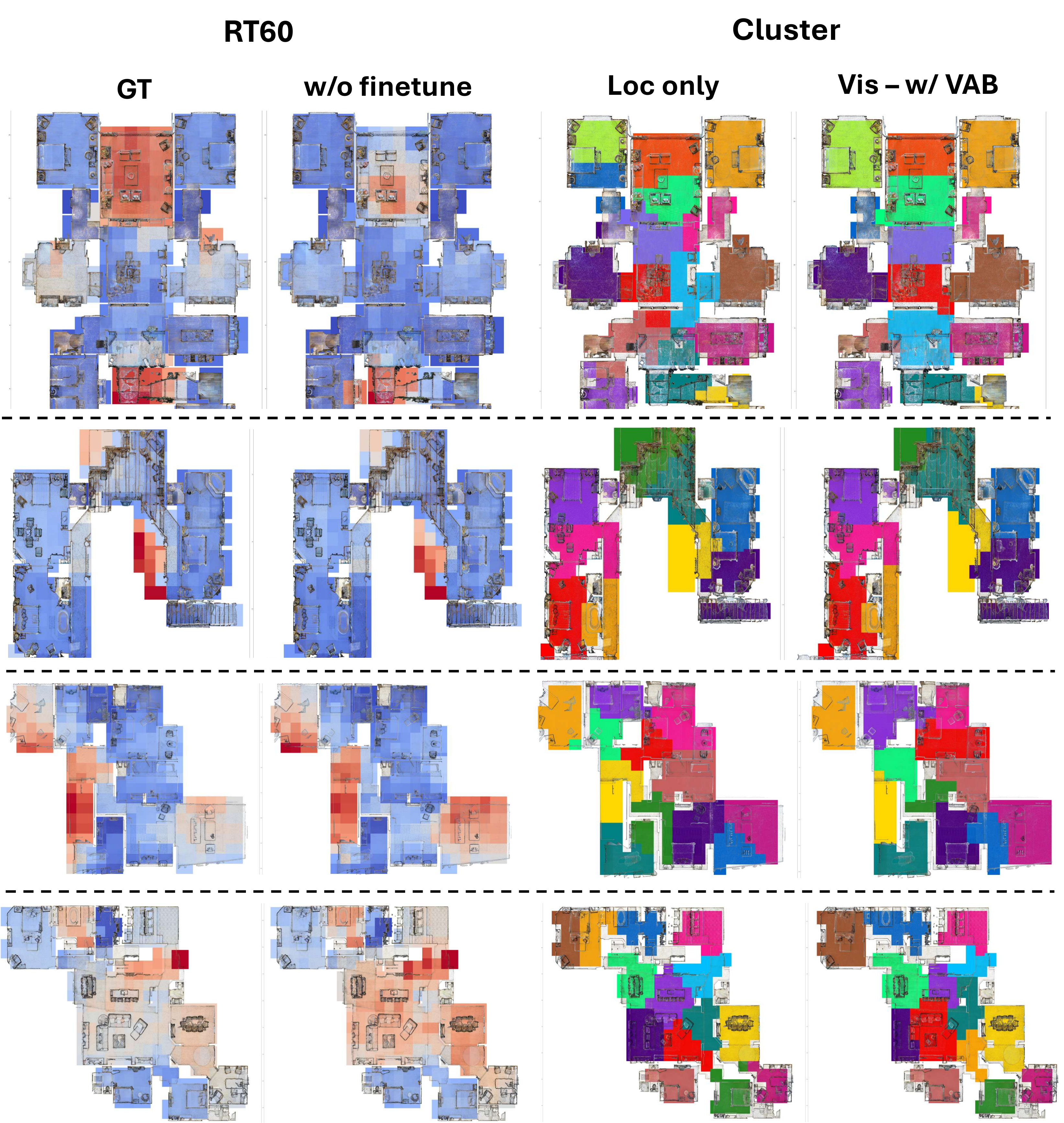}
    \caption{\textbf{Seen Scenes from Soundspace-Ambient Matterport3D Benchmark.} First two columns: \(\mathbf{RT}_{60}\) maps, with warmer colors indicating higher values (longer energy decay). Last two columns: Cluster results comparison, with different colors marking different clusters. Our sampler, \textit{Vis w/VAB}, provides more reliable clusters and the cluster segments better alignment with the \(\mathbf{RT}_{60}\) map.}
    \label{fig:suppl_seen_vis}
\end{minipage}
\end{figure*}

\begin{figure*}[htb]
\small
    \centering
    \includegraphics[width=0.8\linewidth]{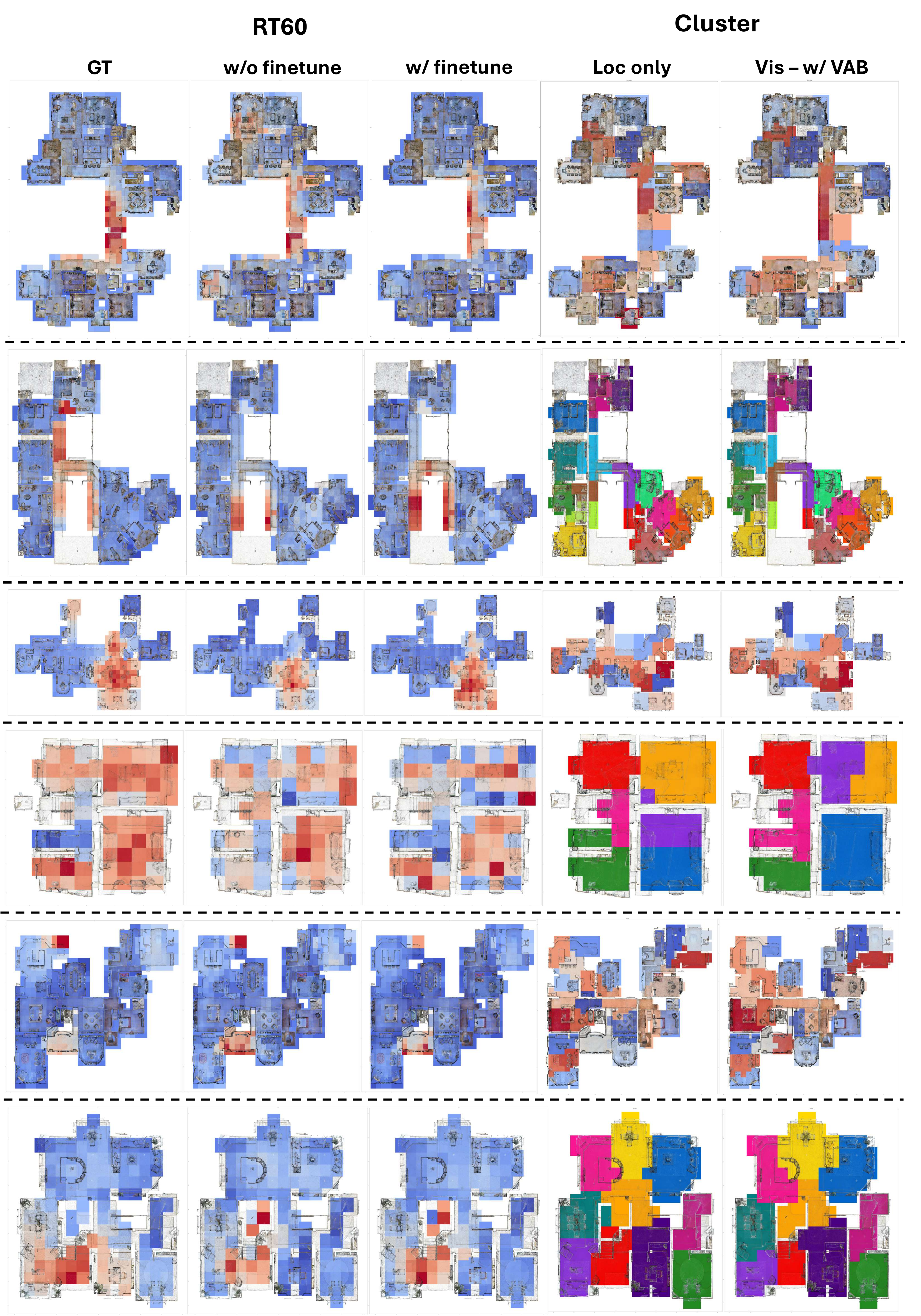}
    \caption{\textbf{Unseen Scenes from Soundspace-Ambient Matterport3D Benchmark.} First three columns: \(\mathbf{RT}_{60}\) maps, warmer colors indicate higher values. \textit{w/ finetune} enhances \(\mathbf{RT}_{60}\) prediction with few-shot finetuning. Last two columns: Cluster results comparison, with colors marking clusters. Our sampler, \textit{Vis w/VAB}, provides more reliable clusters and the cluster segments better align with the \(\mathbf{RT}_{60}\) map.}
    \label{fig:suppl_unseen_vis}
\end{figure*}

We show more visualization exmaples of clustering results via VAB in \cref{fig:suppl_seen_vis} and \cref{fig:suppl_unseen_vis}, covering both seen and unseen scenes, respectively. In both figures, the last two columns display scene clusters in different colors. Our sampler, \textit{Vis w/VAB}, produces cluster segment maps that closely align with \(\mathbf{RT}_{60}\) segments, which effectively highlights obstacles affecting sound propagation. \textit{SoundVista} achieves this by binding visual and acoustic representation through the VAB module, enabling \textit{Vis w/VAB} to identify acoustic regions and key obstacles more effectively than \textit{Loc only}, resulting in more reliable clustering outcomes.

\section{More Details for N2S Real Dataset}
We intentionally partitioned a real room space to create distinct acoustic zones for our N2S benchmark (\cref{sec: n2s_benchmark}). A sound-absorbing divider separates the larger room, while the smaller concrete-walled room is more reverberant than the sound-treated main room. The top view of the geometry of the room is shown as \cref{fig:sim2real_cluster}.
The dataset includes ambient noise from a refrigerator, coffee machine, air vents, and fans; which are challenging to isolate and measure. These add to significant acoustic complexity, although the dataset includes a single scene.

\section{Limitations}
Our method relies on reference recordings, requiring a microphone setup and data collection. These processes can be integrated with existing camera setups for NVS tasks. Additionally, the reliability of our reference sampler may decrease in regions with extremely complex scene layouts. This could be mitigated by incorporating more representative 3D visual descriptions to enhance the VAB module.

\section{Broader Impact}
Our pipeline can produce audio recordings that mimic real recordings from a specific room. However, this capability can lead to the creation of deceptive and misleading media. It is worth noting that, our model doesn't generate new content; instead, it primarily adapts the pre-recorded audio to sound as if it were captured from the target positions. 

\end{document}